\documentclass[sigconf]{acmart}

\usepackage{amssymb}
\usepackage{graphicx}
\usepackage{url}
\usepackage{color}
\usepackage{amsmath}
\usepackage{mathtools}
\graphicspath{ {images/} }
\usepackage{enumitem}

\definecolor{grey}{rgb}{1,0.5,0.5}

\setcitestyle{numbers,sort&compress}
\renewcommand\footnotetextcopyrightpermission[1]{} 

\begin{document}
\title{Characterizing User-to-User Connectivity with RIPE Atlas}

\author{Petros Gigis}
\affiliation{
  \institution{RIPE NCC, Netherlands}
}
\affiliation{
  \institution{FORTH-ICS, Greece}
}
\email{gkigkis@ics.forth.gr}

\author{Vasileios Kotronis}
\affiliation{
  \institution{FORTH-ICS, Greece}}
\email{vkotronis@ics.forth.gr}

\author{Emile Aben}
\affiliation{
  \institution{RIPE NCC, Netherlands}}
\email{emile.aben@ripe.net}

\author{Stephen D. Strowes}
\affiliation{
  \institution{RIPE NCC, Netherlands}}
\email{sds@ripe.net}

\author{Xenofontas Dimitropoulos}
\affiliation{
  \institution{FORTH-ICS, Greece}}
\email{fontas@ics.forth.gr}

\begin{abstract}
Characterizing the interconnectivity of networks at a country level is an interesting but non-trivial task. The IXP Country Jedi~\cite{ixp-country-jedi} is an existing prototype that uses RIPE Atlas probes in order to explore interconnectivity at a country level, taking into account all Autonomous Systems (AS) where RIPE Atlas probes are deployed. In this work, we build upon this basis and specifically focus on ``eyeball'' networks, \emph{i.e.} the user-facing networks with the largest user populations in any given country, and explore to what extent we can provide insights on their interconnectivity. In particular, with a focused user-to-user (and/or user-to-content) version of the IXP Country Jedi we work towards meaningful statistics and comparisons between countries/economies. This is something that a general-purpose probe-to-probe version is not able to capture.
We present our preliminary work on the estimation of RIPE Atlas coverage in eyeball networks, as well as an approach to measure and visualize user interconnectivity with our Eyeball Jedi tool.

\end{abstract}




\maketitle

\section{Introduction} \label{section:intro}

Eyeball networks, \emph{i.e.} the networks that provide Internet access to end-users at the ``last mile'', are interesting for multiple reasons. As opposed to content which can be moved around and hosted anywhere in the network~\cite{nygren2010akamai}, end-users usually access the Internet from a limited physical area, typically the area where they reside or work. Therefore, routing IP packets from end-user to end-user is really determined by how the networks serving the users are connected with each other. While the majority of traffic today is eyeball-to-content (\emph{e.g.} users accessing content ``giants''~\cite{labovitz2010internet} such as Google, Facebook, Akamai, Cloudflare, Netflix), and these are traffic flows that are optimized for latency~\cite{chiu2015we}, user-to-user traffic is important for real-time communications (\emph{e.g.} VoIP or online gaming). For reasons of efficiency---and potentially also security--- local (\emph{e.g.} country-level) user-to-user traffic should \emph{stay local}: paths between users ought thus to traverse the fewest intermediaries possible, especially when the end-users are closely geo-located. 

The RIPE Atlas measurement platform~\cite{ripe-atlas} has probes located inside multiple user networks; these probes can perform active Internet measurements such as traceroutes.
In this work we explore what RIPE Atlas probes can tell us about the interconnectivity of networks which serve the majority of users in a given country. The following interesting questions arise in this context:

\textbf{(Q1)} How many of the eyeball networks within a country contain RIPE Atlas probes, and are thus measurable?

\textbf{(Q2)} Do the ``traffic locality'' and ``direct connectivity'' properties actually hold for the majority of users at a country level?

Answering these questions presents a number of challenges:

\textbf{(C1)} What is the most suitable approach to visualize the associated statistics and measurements?

\textbf{(C2)} What is a suitable probe selection strategy in order to infer common connectivity between eyeball networks within a country, while also minimizing measurement costs?

\textbf{(C3)} How can we verify and amortize the influence of IP-to-AS mapping and IP-level geo-location artifacts?

In this short paper we answer questions (Q1) (see Section~\ref{section:ripe-atlas-coverage}) and (Q2) (see Section~\ref{section:eyeball-country-jedi}), focusing on addressing challenge (C1). Approaches to address challenges (C2) and (C3) are discussed as future work in Section~\ref{section:discussion-future}.
Our intent is to understand and characterize aspects of user-to-user connectivity, even in light of limited probe coverage within their eyeball ISPs. We believe that such a characterization can help both network operators and Internet users discover interesting interconnectivity artifacts and issues~\cite{obar2013internet,gupta2014peering} within the countries they operate and live, and act upon them. We build a new tool to achieve this, the \emph{Eyeball Jedi} (see Section~\ref{section:eyeball-country-jedi}).

\vspace{-3mm}
\section{Estimating User Coverage} \label{section:ripe-atlas-coverage}

As a prerequisite to understanding interconnectivity, we first need to be able to measure eyeball networks. Therefore, we need to have the necessary infrastructure deployed within the networks and understand how many users are covered by such deployments~\cite{eyeball-cover-article}.
We thus measure and visualize the population coverage by RIPE Atlas probes per country \emph{systematically in time}, taking into account the deployment of IPv6/IPv4 probes in the most populated eyeball networks. We use the following data sources: (i) RIPE Atlas probe data (including their location information), which are fetched
using the RIPE Atlas API~\cite{ripe-atlas-api},
(ii) Internet users per country data which are based on Internet Live Stats~\cite{livestats}, and~
(iii) user population per AS data, based on APNIC estimates~\cite{apnic-aspop}. The latter are derived from the APNIC IPv6 measurement campaign~\cite{apnic-v6meas}.
On a daily basis we fetch the biggest eyeball networks from APNIC data for all available countries.
Per country we use the inferred eyeball populations to estimate which networks (AS) are the dominant players up to a cumulative fraction of 95\% of the Internet users in that country. As a population coverage threshold \emph{per AS}, we consider a value of 1\% since this allows us to study the country-level eyeball ecosystems at a fine granularity. 
This method typically represents a majority of Internet users, on average covering $\sim$90.5\% of end-users per country, though there are outliers such as Russia with only 29.3\% coverage due to a highly fragmented eyeball ecosystem.
As a next step we use the readily available data from IXP Country Jedi~\cite{ixp-country-jedi} to determine if there were any probes in these networks participating in mesh-measurements during the latest runs of the tool.
Eyeball networks with at least one public probe are considered as covered. Using this methodology we generate the number of Internet users being covered per AS, based on the population percentage of the AS (APNIC data) and the number of Internet users of the AS's country (Internet Live Stats).
These data are made publicly available for all countries in both tabular formats~\cite{eyeball-cover-table} and on color-coded world maps~\cite{eyeball-cover-map}.
We note that the challenge of the RIPE Atlas deployment limitations and missing networks was first presented at~\cite{ripe-cover}. 

\vspace{-4mm}
\section{The Eyeball Jedi}
\label{section:eyeball-country-jedi}

The Eyeball Jedi launches and processes traceroutes monthly, from eyeball to eyeball network, on a country level, for all available countries, after selecting the required probes per network (AS). These networks were identified in Section~\ref{section:ripe-atlas-coverage}. 
Eyeball-to-eyeball AS-paths are extracted from IP-level traceroutes using RIPEstat~\cite{ripestat} IP-to-AS mapping. While for this analysis we consider IPv4 datasets, IPv6 is also supported.
Currently, we use the following probe selection methodology per AS per country: we select the closest and the most distant probe using the capital of a country as a point of reference, similarly to the IXP Country Jedi~\cite{ixp-country-jedi}. This strategy aims to exploit geo-related path diversity effects between eyeballs, using a minimal number of probes. The location information of the probes is provided by RIPE Atlas. Furthermore, to geo-locate the traceroute hops involved in the Eyeball Jedi measurements, we use OpenIPMap~\cite{openipmap}. This is important to infer whether the inter-eyeball traffic remains within or exits a country. Using this approach, and assuming that the probes we select actually capture the diversity of the eyeball interconnection, and accurately represent the local market, we can estimate the percentage of user-to-user connections that stay in (or go out of) the country, as well as whether they are direct or not.
It is future work to assess if this assumption holds in practice.

\begin{figure} [ht]
  \centering
 \includegraphics[width=0.9\columnwidth]{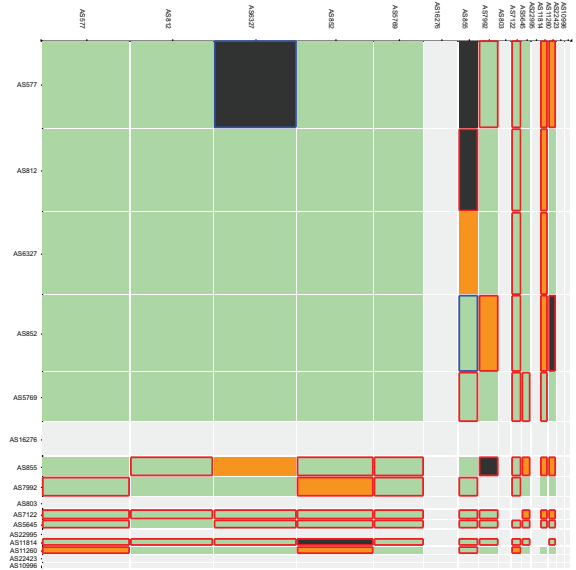}
 	\vspace{-4mm}
	\caption{Snapshot of eyeball-to-eyeball matrix for Canada (generated on 2017-04-01).   Colors are mapped as follows: In-country paths are green, out-of-country orange. Eyeball networks without RIPE Atlas probe coverage are light grey; in-/out-of-country inconsistencies between probes are black.
    Red borders mark indirect AS-level eyeball connections; blue borders mark direct/indirect inconsistencies.}
  \label{fig:ca_latest}
\end{figure}

To visualize our results, we generate a tabular structure called the AS-to-AS matrix. Rows and columns correspond to different eyeball networks (AS), used as sources and destinations respectively. The resulting boxes are sized according to the APNIC estimations of the coverage of Internet users per AS. The colors of the boxes correspond to different types of interconnectivity information, such as out-of-country or in-country, while red borders mark indirect AS-level connections. With such a structure, we can calculate metrics related to the user population that interconnects via (direct or not) paths within or outside a country. The basic metric we use is reflected by the area of the displayed boxes, which corresponds to a product of coverage percentages. By dividing such areas with the total area, we can calculate percentages of user-to-user connections with certain characteristics. An example for Canada is depicted in Fig.~\ref{fig:ca_latest}. First, we found that 16 AS cover the 84.5\% of Internet users in Canada. Second, the cumulative area of user connections, seemingly served via in-country paths, is 47.1\%. Only 9\% are indirect (with $\geq1$ intermediaries) on the AS-level.
3.1\% leave the country. Moreover, 18.1\% suffer from lack of RIPE Atlas probe coverage, and only 3.2\% exhibit inconsistencies w.r.t. achieving consensus on whether traffic actually leaves the country or remains within it. 28.6\% is the rest of the area not examined by the Eyeball Jedi. We note that the asymmetries displayed in the matrix may stem either from inference errors or from interesting ISP policy differentiation per traffic direction, something we plan to investigate further.

\section{Conclusions \& Future Work}
\label{section:discussion-future}

We presented a new tool, the Eyeball Jedi, that can be used to characterize aspects of the interconnectivity between eyeball networks on a country level.
The tool uses probe-to-probe traceroutes from RIPE Atlas and estimated population coverage data to yield useful user-to-user statistics and visualizations. 
As future work, we intend to turn our initial implementation available in~\cite{eyeball-cover-table,eyeball-cover-map,eyeball-country-jedi,eyeball-jedi-github} into a fully-fledged publicly available tool. In addition, we aim at gaining visibility in
the eyeball-to-neighbor-country-eyeball traffic.
W.r.t.~challenge (C2), we plan to investigate more sophisticated probe grouping and selection strategies.
Addressing challenge (C3) will require attaining proper ground truth for validation of the inferred characterization (\emph{e.g.} via crowd-sourced verification).

\newpage
\section{Acknowledgements}
This work has been partly funded by the European Research Council Grant Agreement no. 338402.

\bibliographystyle{acm}
\bibliography{biblio}

\end{document}